# Experimental Demonstration of 503.61-Gbit/s DMT over 10-km 7-Core Fiber with 1.5-μm SM-VCSEL for Optical Interconnects

Lu Zhang[(1,2)], Joris Van Kerrebrouck[(3)], Oskars Ozolins[(4)], Rui Lin[(1)], Xiaodan Pang[(1,4,8)], Aleksejs Udalcovs[(4)], Siliva Spiga[(6)], Markus C. Amann[(6)], Lin Gan[(5)], Ming Tang[(5)], Songnian Fu[(5)], Richard Schatz[(1)], Gunnar Jacobsen[(4)], Sergei Popov[(1)], Deming Liu[(5)], Weijun Tong[(7)], Guy Torfs[(3)], Johan Bauwelinck[(3)], Xin Yin[(3)], Shilin Xiao*[(2)], Jiajia Chen*[(1)]

[(1)] KTH Royal Institute of Technology, Kista, Sweden, *jiajiac@kth.se*
[(2)] State Key Laboratory of Advanced Optical Communication System and Networks, Shanghai Jiao Tong University, Shanghai, China, *slxiao@sjtu.edu.cn*
[(3)] Department of Information Technology (INTEC) – IDLab, University of Ghent – IMEC, Gent, Belgium
[(4)] Networking and Transmission Laboratory, RISE AB, Kista, Sweden
[(5)] Huazhong University of Science and Technology, Wuhan, China
[(6)] Walter Schottky Institut, Technische Universität München, München, Germany
[(7)] Yangtze Optical fiber and Cable Joint Stock Limited Company, Wuhan, China
[(8)] Infinera, Fredsborgsgatan 24, 117 43 Stockholm, Sweden

**Abstract** *We experimentally demonstrate a net-rate 503.61-Gbit/s discrete multitone (DMT) transmission over 10-km 7-core fiber with 1.5-μm single mode VCSEL, where low-complexity kernel-recursive-least-squares algorithm is employed for nonlinear channel equalization.*

**Introduction**

To fulfill the capacity demand of datacenter networks while keeping low cost, intensity modulation and direct detection (IM/DD) systems with links of 200-GbE, 400-GbE and beyond have been widely investigated[1]. Meanwhile, vertical-cavity surface-emitting lasers (VCSELs) have merits such as low cost, low power and ease of integration into arrays, which is promising for datacenter interconnects. There are many works on short-wavelength (e.g. 850nm) multimode VCSELs and multimode fibers (MMF) based short-reach interconnects with advanced modulation formats (e.g. pulse amplitude modulation PAM[2], discrete multi-tone DMT[3], carrier-less amplitude and phase modulation CAP[4]). However, transmission distance via MMF is limited up to a few hundred meters due to mode dispersion. To the best of our knowledge, the longest transmission distance is 0.55-km MMF with 106-Gbit/s net-rate per channel[3] at hard-decision forward error correction (HD-FEC) limit[5].

To increase the reach and decrease the loss in silicon-based fiber link, long-wavelength single-mode (SM) VCSELs have gained many interests[6-11]. In our previous work[11], we have achieved 726-Gbit/s DMT transmission over 2.5-km multi-core fiber (MCF), using recursive-least-squares (RLS) based Volterra filtering scheme for nonlinear equalization. However, the complexity of Volterra filtering algorithm might be too high for optical interconnects that are cost- and latency-sensitive.

In this paper, we experimentally demonstrate DMT transmission over 10-km 7-core fiber with 1.5-μm SM-VCSEL. Low-complexity kernel RLS (KRLS) algorithm is introduced for nonlinear equalization instead of conventional Volterra-RLS scheme. With KRLS, we successfully generate 103.76-Gbit/s DMT signal for optical back-to-back (OBtB) transmission and achieve total net-rate of 503.61-Gbit/s transmission over 7-core-fiber at 7% overhead HD-FEC limit with BER of 3.8e-3.

**Experimental setup**

The experimental setup is shown in Fig. 1, The digital signal processing (DSP) flow is also included in Fig. 1. The DMT signals are generated offline and loaded into a 92-GSa/s digital to analog converter (DAC). The length of the inverse fast Fourier transform (IFFT) points and cyclic prefix length of the DMT signal are set to 1024 and 16, respectively.

In the experiment, the VCSEL die is electrically driven via a 100-μm GSG 50-GHz probe. The light generated in the VCSEL is coupled into a single-mode lens fiber. There is no temperature controller (TEC) used in the setup. The VCSEL bias current is set to 7.8-mA. The measured central wavelength of the probed

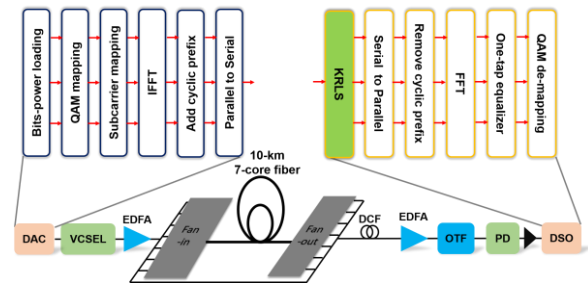

**Fig. 1:** Experimental setup

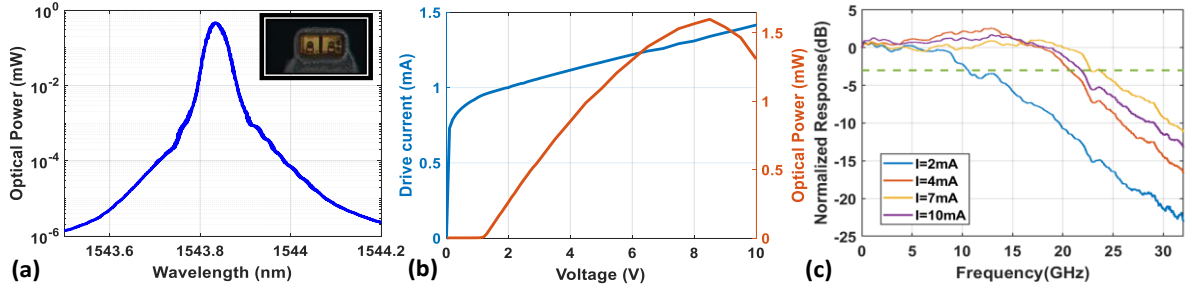

**Fig. 2:** (a) optical spectrum of VCSEL, (b) The P-I-V curves and (c) small-signal $S_{21}$ responses of VCSEL.

VCSEL is 1543.2-nm and the captured output optical power is 1-dBm. The optical spectrum (measured by the optical spectrum analyzer with 0.037-nm bandwidth resolution), P-I-V curve and the small signal $S_{21}$ response of the probed VCSEL are shown in Fig. 2(a), Fig. 2(b) and Fig. 2(c), respectively. The maximum 3-dB bandwidth is about 22-GHz. The signal is split in a fan-in module of the 10-km 7-core MCF, using 1:8 optical coupler. An Erbium-doped fiber amplifier (EDFA) with 14.8-dBm output power is used before fan-in module to compensate for extra loss from de-correlation. The optical delay lines are used to de-correlate the signals to emulate a practical system using seven independently modulated VCSELs. After the 10-km 7-core MCF (inter-core crosstalk: -45dB/100km), signals are detected individually after a fan-out module. A dispersion compensation fiber (DCF, -159 ps/nm) is employed after the MCF. The signal is amplified by a pre-amplifier EDFA and an optical tunable filter (OTF) is utilized to filter out the amplified spontaneous emission (ASE) noise. A 90-GHz PIN photo-detector (PD) is used at the receiver. A variable optical attenuator (VOA) is used before the PD for bit error rate (BER) curves. The electrical signal after direct-detection is amplified by a 65-GHz linear electrical amplifier and captured by a 160-GSa/s digital storage oscilloscope (DSO). After DSO, the captured signal is processed offline.

KRLS is used for nonlinear equalization, which has the same multi-tap structure as linear RLS filtering scheme. For RLS, it aims at minimizing the sum of squared estimation errors up to the current iteration $i$. In RLS training, the input is the received signal $u(i)$ and the target is to minimize the error between the desired signal $d(i)$ and the received signal $u(i)$ by optimizing the filter coefficient $w(i)$, which is expressed as: $\min_w \sum_{j=1}^{i} |d(j) - u(j)^T w|^2 + \lambda \|w\|^2$, where $\lambda$ is the regularization factor. For KRLS, the input is the mapping of received signals in the kernel Hilbert space[12], denoted as $\varphi(u(i))$, and the target is $\min_w \sum_{j=1}^{i} |d(j) - \varphi(u(j))^T w|^2 + \lambda \|w\|^2$. The mapping $\varphi(.)$ is the feature mapping and $\varphi(u(i))$ is the transformed feature vector in the kernel Hilbert feature space. Then, we can formulate the classic RLS linear adaptive filters in the kernel Hilbert space and iteratively solve a convex nonlinear problem. The nonlinear filtering method in the kernel Hilbert space has the same complexity order as the linear RLS.

By introducing $D(i) = [d(1), d(2), ..., d(i)]^T$ and $\Phi(i) = [\varphi(u(1)), \varphi(u(2)), ..., \varphi(u(i))]^T$, the optimal KRLS coefficient can be obtained as:

$$\begin{aligned} \omega(i) &= \Phi(i)[\lambda I + \Phi(i)^T \Phi(i)]^{-1} D(i) \\ &= \Phi(i) a(i) \\ &= \Phi(i) P(i) d(i) \end{aligned} \quad (1)$$

By introducing:

$$\begin{aligned} h(i) &= \Phi(i-1)^T \varphi(u(i)) \\ &= \{\kappa(u(i), u(1)), ... \kappa(u(i), u(i-1))\}^T \\ z(i) &= P(i-1) h(i) \quad (2) \\ r(i) &= \lambda + \varphi(u(i))^T \varphi(u(i)) - z(i)^T h(i), \end{aligned}$$

the updating function of $P(i)$ and coefficient function $a(i)$ can be derived out. Here, $\kappa(u(i), u)$ is the basis kernel function in the kernel Hilbert space, where Gaussian kernel function[13] is used in this paper. The DSP flow of KRLS for channel equalization is shown in Fig. 3.

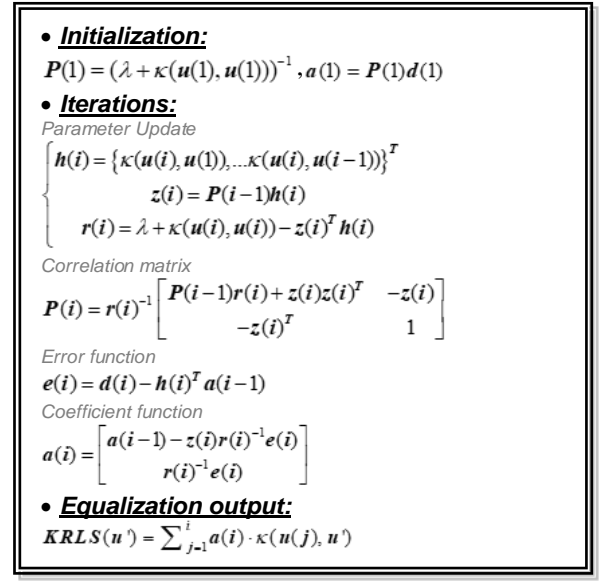

**Fig. 3:** DSP flow of KRLS for channel equalization

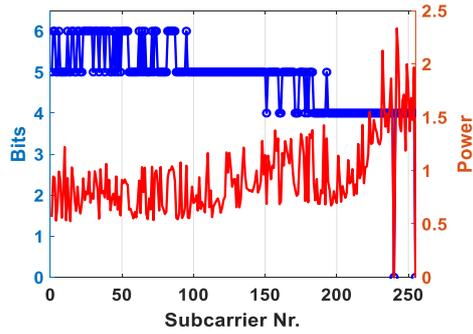

**Fig. 4:** Bits and power loading profiles at OBtB case

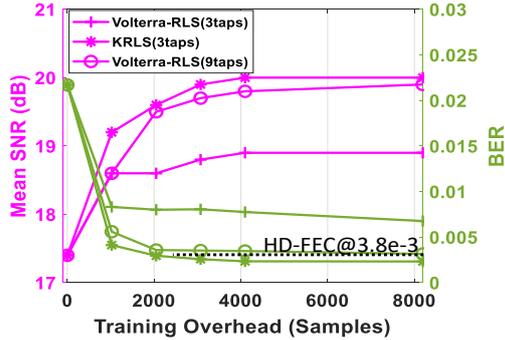

**Fig. 5:** Volterra-RLS versus KRLS in terms of training overhead and equalization taps

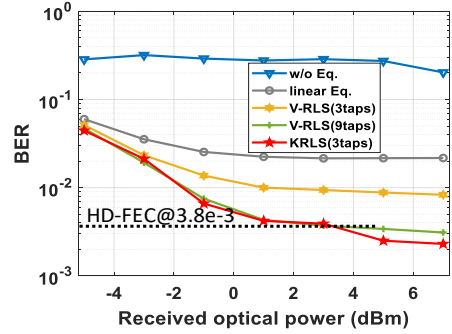

**Fig. 6:** BER performance at OBtB case

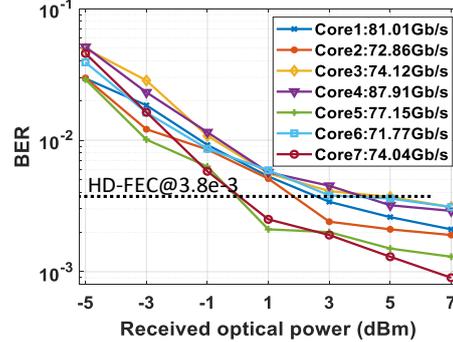

**Fig. 7:** BER performance after 10-km MCF with KRLS

## Experimental results

In the OBtB case, 254 subcarriers are loaded. Since OBtB case is a low-pass channel, bit-power loading can help with improving the system capacity and spectrum efficiency. The bit-power loading profiles of the OBtB case is shown in Fig. 4. The quadrature amplitude modulation (QAM) orders vary from 64QAM to 16QAM. In 10-km MCF case where 210 subcarriers are loaded, bit-power loading is also employed.

The KRLS is firstly used to optimize OBtB case, which is shown in Fig. 5. The original signal-to-noise-ratio (SNR) and BER is measured with linear one-tap channel equalization. Here, Volterra-RLS considers both the 2$^{nd}$ and 3$^{rd}$ order nonlinear distortions. With 3-tap Volterra-RLS, the SNR is improved by 1.5-dB with training overhead larger than 4096 samples and the BER is still larger than the HD-FEC limit. Volterra-RLS reaches the HD-FEC limit with 9-tap 2$^{nd}$ and 3$^{rd}$ order equalization and more than 4096 samples. In contrast, KRLS reaches BER lower than the HD-FEC limit with only 3 equalizer taps and 4096 training samples and improves SNR by 2.6-dB.

The BER performance versus the received optical power at OBtB is shown in Fig. 6. The achieved line rate at OBtB is 103.76-Gbit/s, the performance with nonlinear equalization is significantly improved compared to the ones without equalization and with linear equalization.

The BER performance of 10-km 7-core fiber transmission is shown in Fig. 7. The achieved line-rates for different cores are 81.01-Gbit/s, 72.86-Gbit/s, 74.12-Gbit/s, 87.91-Gbit/s, 77.15-Gbit/s, 71.77-Gbit/s, 74.04-Gbit/s, respectively. The total system capacity with 10-km 7-core fiber is 538.86-Gbit/s (net-rate 503.61-Gbit/s).

## Conclusions

We have experimentally demonstrated a net-rate 503.61-Gbit/s DMT transmission over 10-km 7-core fiber with 1.5-µm SM-VCSEL, where 3-tap KRLS algorithm is employed realizing low-complexity nonlinear equalization with BER lower than the HD-FEC limit.

## Acknowledgements

This work was partly supported by the Natural Science Foundation of China (#61331010, 61722108, 61775137, 61671212), H2020 5GPPP 5G-PHOS research program (ref.761989), European Commission through the FP7 project MIRAGE (ref.318228), Swedish Research Council (VR), the Swedish Foundation for Strategic Research (SSF), Göran Gustafsson Foundation, Swedish ICT-TNG, EU H2020 MCSA-IF Project NEWMAN (#752826), the Celtic-Plus sub-project SENDATE-EXTEND & SENDATE FICUS.

## References

[1] K. Zhong, et al., JLT, vol. 36, no. 2, pp. 377-400, 2018.
[2] F. Karinou et al., in Proc. ECOC, M2C.2, 2016
[3] C. Kottke et al., in Proc. OFC, W4I.7, 2017.
[4] W. Bo et al., in Proc. ECOC, Th2P2SC4, 2016.
[5] ITU -T Recommendation G.975.1, 2004, Appendix I.9.
[6] S. Spiga et al., JLT, vol. 35, no. 4, pp. 727-733, 2017.
[7] X. Yin et al., in Proc. OI, pp. 33-34, 2017.
[8] C. Xie et al., in Proc. OFC, Tu2H.2, 2015.
[9] N. Eiselt et al., JLT, vol. 35, no. 8, pp. 1342-1349, 2017.
[10] X. Pang et al., in Proc. OFC, M1I. 4, 2018.
[11] J. Van Kerrebrouck et al., in Proc. OFC, M1I. 2, 2018.
[12] E. Yaakov et al., IEEE Trans. on Signal Processing, vol. 52, no. 8 pp. 2275-2285, 2004.
[13] W. Liu et al., vol. 57. John Wiley & Sons, 2011.